\documentstyle[aps,prl,multicol,epsf]{revtex}
\begin{document}
\title{\bf Theoretical Results for Sandpile Models of SOC with Multiple
Topplings}

\author{Maya Paczuski$^{1}$ and Kevin E. Bassler$^2$}
\address{Department of Mathematics, Huxley Building, Imperial College
of Science, Technology, and Medicine, London UK SW7 2BZ \\
$^2$ Department of Physics, University of Houston, Houston TX
77204-5506 \\ }
\date{\today}

\maketitle 

\begin{abstract}
 
We study a directed stochastic sandpile model of Self-Organized
Criticality, which exhibits recurrent, multiple topplings, putting it in a
separate universality class from the exactly solved model of Dhar and
Ramaswamy.  We show that in the steady-state all stable states are
equally likely.  Then we explicitly derive a discrete
dynamical equation for avalanches on the lattice.  By coarse-graining
we arrive at a continuous Langevin equation for the propagation of avalanches
and calculate all the critical exponents characterizing
them.  The avalanche equation is similar to
the Edwards-Wilkinson equation, but with a noise amplitude that is a
threshold function of the local avalanche activity, or interface
height, leading to a stable absorbing state when the avalanche dies.
It represents a new type of absorbing state phase transition.
\end{abstract}

{PACS numbers: 05.65.+b, 87.23.Ge, 87.23.Kg}

\begin{multicols}{2}
\section{Introduction}

Sandpile models of stick-slip dynamics have received considerable
attention as canonical models of self-organized criticality (SOC) \cite{soc}.
SOC refers to the widespread tendency of many extended, dissipative
dynamical systems to evolve inevitably towards a complex state with
power-law correlations in space and time: a ``critical'' state.  Of
course, a critical state is only one possible example of complex
phenomena that can emerge in large, self-organizing systems composed
of many strongly interacting parts.  No doubt there are other types of
complex states that have not yet been so well characterized
mathematically, e.g. for example in networks \cite{nk}.  From this viewpoint,
the phenomena of SOC itself is a prototype for how complexity emerges
in nature without fine tuning parameters.  In spite of the gross
simplicity of various cellular models that have been introduced, and
hundreds if not thousands of numerical studies of SOC, only minimal
analytic understanding has been achieved.

In fact, a survey of analytic works on sandpile models of SOC is
exceedingly short.  The model of SOC introduced by Bak, Tang, and
Weisenfeld (BTW) \cite{btw} has yielded to some analytic treatment
associated with its abelian properties, primarily due to the work of
Dhar and collaborators \cite{dharreview}.  The scaling properties of
waves, where each site only topples, or releases grains, once has been
understood by Priezzhev and collaborators \cite{prieezhev,KLGP}.
Nevertheless, the large scale properties of avalanches, where each
site can topple many times in response to a single grain being added
to the system, remain unsolved and the numerical situation
controversial \cite{KLGP,one,two,three}.  The same is true for the
Zhang model where some limited progress has been made
using methods from dynamical systems theory \cite{cessac}.  Recurrent,
multiple topplings within an avalanche also appear in most other unsolved
sandpile models, such as the stochastic Manna model \cite{manna}, the
universality class \cite{universal} represented by the Oslo rice pile
model \cite{oslo}, cellular models of vortex dynamics
\cite{vortices}, as well as   one-dimensional trough
models exhibiting multiscaling \cite{kadanoff}.  The
difficulties preventing progress in solving any of these simplified
models in particular, or finding general analytic tools for granular
systems exhibiting SOC, appear to be related, in part, to the
existence of recurrent topplings.

This statement is further supported by the following facts: Dhar and
 Ramaswamy (DR) \cite{dr} introduced a directed version of the BTW
 model, and solved for the avalanche distribution and many other
 properties exactly.  In the DR model, it can be rigorously proven
 that no multiple topplings occur.  (Consequently, the elegant DR
 solution, as it has been conceived thus far, does not address the
 full complexity of discrete or granular models of SOC.)  The fixed
 scale transformation method of Pietronero and collaborators
 \cite{piet} also explicitly ignores the presence of multiple
 topplings.  One consequence of this fact is that this method puts the
 stochastic Manna model and the BTW model into the same universality
 class, which is not consistent with most numerical works
 \cite{one,biham} (except \cite{two}), including those measuring
 unequivocal differences in aging behaviors \cite{aging}.  Multiple
 topplings, where the activity can return an arbitrarily
large number of times, do not appear in any mean field description
 \cite{mean}, since in high enough dimensions, the avalanche activity
 is not recurrent, or able to return more than
a finite number of times \cite{bethe}, at any site.
 Multiple topplings are a fluctuation
 effect associated with self-intersections of the avalanche cluster in
 space and time \cite{scaling}, and as such are relevant below some
upper critical dimension.

Certainly, the intricacies associated with multiple topplings are not
the only ones that present themselves in attempting an analytic
treatment of granular models of SOC.  For example, the fact that the
dissipation process is confined to the boundary, which forces the
system to self-organize, is an important and subtle point because the
boundary cannot be scaled out in the limit of large system sizes as is
usually done in statistical physics.  In principle, the boundary is
always important, because the incoming sand grains must be transported
to it, no matter how large the system size.  The broken translational
invariance associated with the boundary often leads to long range
boundary effects in the metastable states (see, for example,
\cite{middleton,drossel}).  It might be useful to pry these
complications apart, treating one issue at a time.  Here we focus on
the problem of recurrent or multiple topplings, and seek a model which
does not present other difficulties.

Recently, Pastor-Satorras and Vespignani \cite{pv} have studied
numerically a stochastic directed sandpile model (SDM), which is a
stochastic version of the exactly solvable model introduced by DR.
This stochastic model is simpler and presumably unrelated to the
directed models introduced and studied by Tadi\'c and collaborators
\cite{tadic}.  Pastor-Satorras and Vespignani demonstrated by
numerical simulations that the model exhibits multiple topplings which
changes the universality class, making it distinct from the DR model.
This was accomplished by numerically measuring and comparing various
critical exponents characterizing the avalanches.  Its close relation
to the DR model, which has an exact solution, suggests to us an
analytic study.

\subsection{Summary}
We proceed with an analysis of the SDM as follows: First we define the
DR model and the SDM.  For pedagogical reasons, in Section III, we
review the proof that the critical state of the DR model is the set of
all stable states with equal probability.  We also review some
necessary parts of Dhar's construction of an operator algebra for
stochastic models, which can be represented as deterministic
models with a quenched array of random numbers.
  Combining these two works, we then show that for
the SDM the critical state is also the set of all stable states with
equal probability, described by a product measure. Using this fact, we
show in Section IV that the SDM can be recast as a generalized
branching process propagating in an uncorrelated environment, enabling
a study of the infinite system.  By carefully analyzing the
microscopic dynamics of this process on the lattice, we explicitly derive a
discrete dynamical equation for the propagation of flowing grains in
avalanches.  In Section V,
coarse graining this discrete equation gives a continuum equation for
avalanches that should describe the large scale properties of any
microscopic model with the same symmetry, conservation of grains, and
stochastic effects. 

 Notably, our equation is similar to the
Edwards-Wilkinson (EW) equation \cite{ew} except that the amplitude of the
nonconservative noise is a Heaviside (theta) function of the local
activity.  {\it Crucially, the noise amplitude is a threshold
function, rather than being a constant, such as the temperature.}  
The height of the interface represents the number of topplings in
an avalanche.  The
steady state that is eventually reached in the limit of large times is
always the state of no activity where the height of the interface is
zero everywhere and the avalanche has died.  Thus the equation
describing avalanche dynamics corresponds to an absorbing state phase
transition where the the transient state is governed by the EW
equation in the region where it survives.  Section V also describes an
analysis of this nonlinear equation.  We extract all the (nontrivial)
critical exponents for avalanches, i.e. in $d=1$, $D=7/4$,
$\tau=10/7$, $z=2$, $\tau_t=D=7/4$ distinct from the DR model.  For $d
\geq 2$, where multiple topplings are not relevant, the critical exponents
are the same as in the DR model.  All of these results agree perfectly
with previous numerical works
\cite{pv,vazquez}.  We also write down the Fokker-Planck
equation for the probability distribution of the number of topplings
at each site in an avalanche, although we do not solve it.  Finally,
we conclude with a brief comment on some possibilities for future analytical
work on absorbing state phase transitions and granular models of SOC.

\section{Definition of Directed Models}

Consider a two dimensional square lattice as shown in Fig. ~1.  The
direction of propagation is labeled by $t$, with $0\leq t < T$.
The transverse direction is labeled by $x$, with periodic boundary
conditions. Only sites with $(x+t)$ even are on the lattice, so that $x$ 
is a positive integer modulo $2X$, and the lattice has a total of
$TX$ sites.  On each site,
an integer variable $z(x,t)$ is assigned.  The $i$'th grain is added
to a randomly chosen site $x_i$ on the top row $t=0$. There
$z(x_i,0)\rightarrow z(x_i,0) +1$.  When any site  acquires a height
greater than $z_c=1$ it topples, i.e. $z(x,t) \rightarrow z(x,t) -2$
for $z(x,t)>z_c$.

\begin{figure}
\narrowtext
\epsfxsize=3.0truein
\epsffile{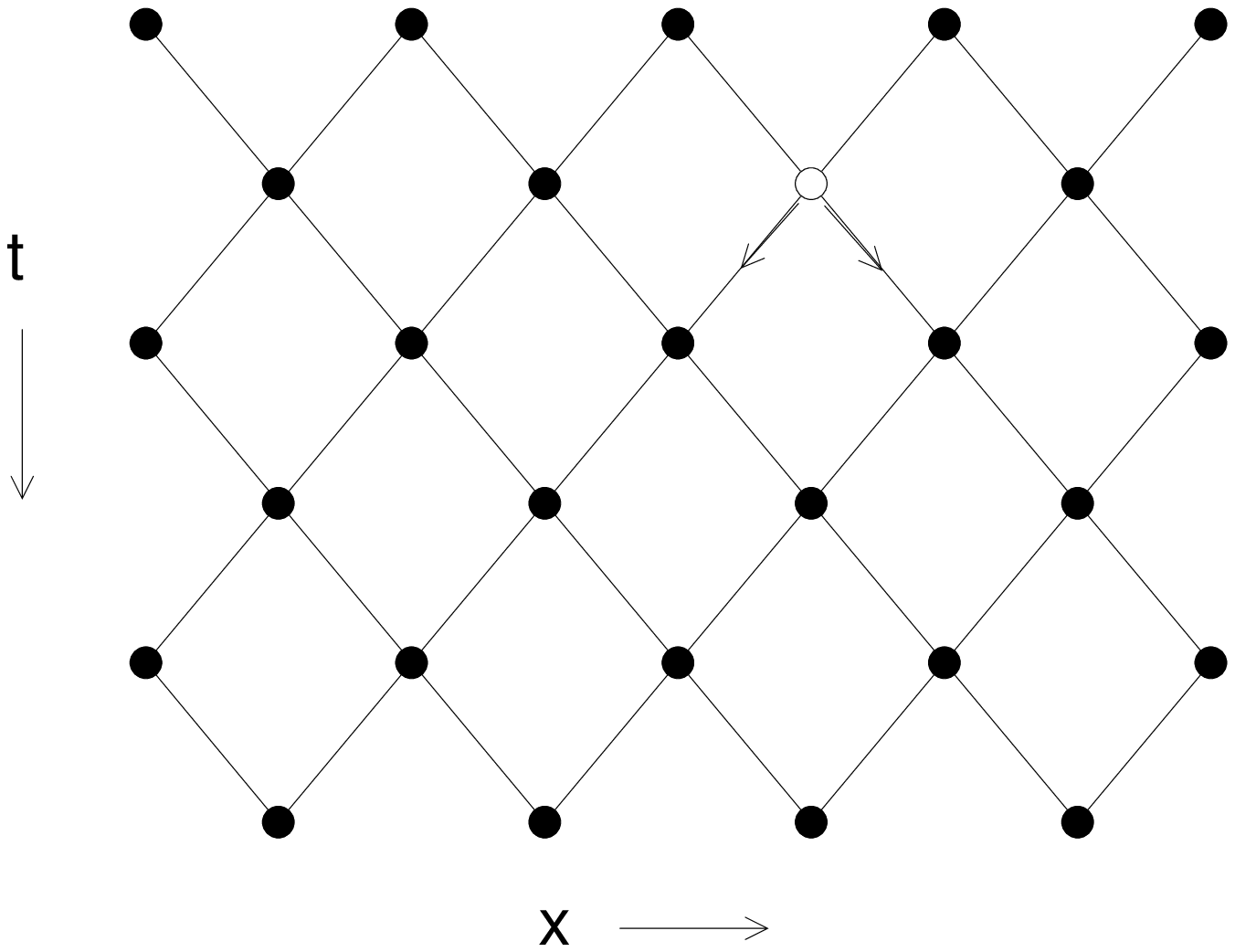}
\caption{
 {\bf Directed Sandpile Models} Grains from active sites in row $t$
topple onto sites in row $t+1$. For example, grains from the site
indicated by the open circle topple only onto the nearest neighbor
sites indicated by the arrows. }
\end{figure}

The two models differ with respect to the transmission of grains out of
a toppling site.  In the DR model, one grain is transferred to the left
downstream neighbor and one grain to the right so the toppling rule is
for $ z(x,t)>z_c$
\begin{eqnarray} 
z(x,t)& \rightarrow & z(x,t) -2 \nonumber \\
z(x-1,t+1)& \rightarrow &  z(x-1,t+1) + 1 \nonumber \\
z(x+1,t+1)& \rightarrow & z(x+1,t+1) + 1  \,\, . \nonumber
\end{eqnarray}
For the SDM, on the other hand, each grain from a toppling site is
given equal probability to go to any downstream nearest neighbor.  In
this case, when the site $(x,t)$ topples,
$$
z(x,t) \rightarrow z(x,t) - 2
$$
and
\begin{eqnarray}
z(x-1,t+1)& \rightarrow & z(x-1,t+1) + 1 \nonumber \\
z(x+1,t+1)& \rightarrow & z(x+1,t+1) + 1  \,\,  \nonumber
\end{eqnarray}
with probability 1/2, or
\begin{eqnarray}
z(x-1,t+1)& \rightarrow & z(x-1,t+1) + 2 \nonumber \\
z(x+1,t+1)& \rightarrow & z(x+1,t+1)   \,\,  \nonumber
\end{eqnarray}
with probability 1/4, or
\begin{eqnarray}
z(x-1,t+1)& \rightarrow & z(x-1,t+1) \nonumber \\
z(x+1,t+1)& \rightarrow & z(x+1,t+1) + 2  \,\, . \nonumber
\end{eqnarray}
with probability 1/4.  Thus, the SDM is
a directed version of the model introduced by Manna.  

In both
directed models, grains are conserved during each toppling event.
This is true except at the open boundary $t=T$ where toppling sites
simply discharge their grains out of the system.  Sites are relaxed
according to a parallel update until there are no more unstable sites,
and the properties of the resulting avalanche are recorded.  Then a
new avalanche is initiated by adding a single grain to a randomly
chosen site on the top row, $t=0$.  An avalanche can be characterized by its
longitudinal extent, $t_c$, the largest $t$ row affected, its width,
$x_c$, the largest transverse distance from the avalanche origin to any
site affected by the avalanche, its area, $a$, the total number of
sites affected, its size, $s$, the total number of toppling events, and
the maximum number of topplings at a site, $n_c$.

It is straightforward to generalize this definition to higher
dimensions, with the number of directions transverse to the direction
of propagation being $d$.  In this case $z_c=2d-1$.  At a toppling
site $z \rightarrow z - z_c -1$.  In the DR case each downstream
neighbor receives exactly one grain.  In the stochastic case, each
downstream neighbor has equal probability $1/2d$ to receive each
grain.  For simplicity of notation and concepts we will focus our
discussion on the case $d=1$ unless otherwise noted.

\section{States on the Attractor}

For both directed models, any configuration satisfying $0 \leq z(x,t)
\leq z_c$ for all $(x,t)$ is stable.  The total number of such
configurations is $(z_c+1)^{TX}$.  For clarity, we now review the argument
 showing that in the steady state, all such stable states are
equally likely in the DR model.

\subsection{Review of Some Exact Results by Dhar and Ramaswamy}

Let $C_0$ be a starting configuration with the $i$'th particle added
at site $x_i$, resulting in the new stable
configuration $C_{i}$.  Then $C_{i}$
is uniquely determined by the dynamics given $x_i$ and $C_{i-1}$.  The
crucial point is that this
dynamics is invertible.  On the top row $C_i$ differs from $C_{i-1}$
only at the site $x_i$, with $z(x_i,0)$ in $C_i$ being more than its value in
$C_{i-1}$ by one (mod2).  Other rows in $C_{i-1}$ are the same as in
$C_i$ if there was no toppling at $(x_i,0)$; otherwise the $z$'s in
the first row, $t=1$, in $C_{i-1}$ are the same as in $C_i$, except
at the two downstream neighbors, $(x_i-1,1)$ and $(x_i +1,1)$ of $(x_i,0)$
whose heights are less by one (mod2) than their values in $C_i$.  This
obviously continues for subsequent rows.  Thus given $C_i$ and $x_i$
we can uniquely determine $C_{i-1}$.  

For a given $C_i$, there are precisely $X$ distinct choices of
$C_{i-1}$ and $C_{i+1}$ corresponding to $X$ distinct, possible
choices of $x_i$ and $x_{i+1}$.  The master equation for the evolution
of probabilities of configurations, is
\begin{equation}
dP(C)/dt = -\sum_{C'}T_{C'C}P(C) +
\sum_{C'}T_{CC'}P(C') \quad . 
\end{equation}
Since there are $X$ distinct choices for the $C'$ into $C$ and also
for the $C'$ out of $C$, each having probability $1/X$, the
probability distribution $P(C_0=a)= cons$, \ \ independent of $a$, is
invariant in time.  Thus the probability distribution of
states on the attractor is
a product measure, with each site independently occupied with one particle
with probability 1/2, otherwise being empty.

 In a recent work, Dhar \cite{ADPM} has shown that the stochastic Manna model 
also
exhibits the Abelian property and is a special case of the Abelian
Distributed Processors Model.  Correspondingly some of the the analytic
techniques of the BTW model also apply to the stochastic Manna model.
It is only necessary to realize that for the stochastic models,
instead of associating probabilities with each toppling, we can assign
to each site an infinite stack of random numbers, uniformly
distributed between zero and one, say.  The quenched random numbers in
each site's stack then determine the allocation of grains during each
toppling event.  Thus, the $q$'th random number at (x,t) determines at
the $q$'th toppling of that site where the grains will go.  There is a
one-to-one correspondence between any realization of the dynamics of
the stochastic model, and the dynamics of a deterministic system with
a random array (chosen appropriately to model the probability
distribution of grain allocation), under the same condition of
particle additions. 

If we specify the height configuration of the sandpile as well as the
infinite stack of random numbers at each site, Dhar shows that the
model is also Abelian.  It is easy to check that given any unstable
configuration with two or more unstable sites, we get the same
configuration by toppling at an unstable site $i$, and then at unstable
site $i'$, as we would get if we first toppled at $i'$ and then at
$i$, if the same list of random numbers in the array is provided.
Iterating this until a metastable state is reached proves the Abelian
property of the model.

\subsection{New Results}

The directed stochastic model is also equivalent to a deterministic
directed model with an infinite stack of quenched random numbers at
each site.  Since the latter model is Abelian we can choose to relax each row,
one site at a time, until it is completely stable, before going on to
the next higher row.  In this case, it is easy to see that the 
deterministic model with quenched random numbers shares the
same property of invertibility as the DR model \cite{comment}.

Let $C_0$ be a starting configuration and $R(x,t,q)$ be the infinite
array of random numbers, with the initial pointers $q_0(x,t)=0$ for all
entries $(x,t)$.  The pointers $q$ in the array $R$ will move as
the sequence of topplings proceeds.  The $i$'th particle being added at site
$x_i$ and the current pointers $q_{i-1}(x,t)$ in the array $R$ known,
this results in a new stable configuration $C_i$, and a new set of
pointers $q_i(x,t)$ in the fixed array $R$.  Invertibility
follows.  In this case we are given the current configuration $C_i$,
the current set of pointers $q_{i}(x,t)$ in the fixed array $R$ and
$x_i$.  In order to prove invertibility we must determine both
$C_{i-1}$ and $q_{i-1}(x,t)$.

On the top row $C_{i-1}$ differs from $C_i$ only at the site $x_i$,
with $z(x_i,0)$ in $C_{i-1}$ being less than its value in $C_i$ by
1(mod2). If $z(x_i,0)=1$ in $C_i$, then no toppling occurred and
$C_{i-1}$ is the same as $C_i$ at all other sites; also the set of
pointers $\{q_{i-1}=q_i\}$.  If $z(x_i,0)=0$ then one toppling occurred
at that site.  We locate the pointer $q_i(x_i,0)$ and move it back one
step in the stack $R(x_i,0,q_i(x_i,0) )$ giving $q_{i-1}(x_i,0)=q_i(x_i,0) -1$.
This pointer now points to a number that tells us where the two grains
were placed.  The heights at the sites in the second row $t=1$ in configuration
$C_{i-1}$ are the same as those in $C_i$ except at the forward
neighbors from $x_i$ that received a grain according to
$R(x_i,0,q_{i-1}(x_i,0))$.  If both sites received a grain then we
apply the same procedure to those sites as we applied to $(x_i,0)$.
If one site receives two grains then that site must have toppled once.  Its
height in the previous configuration is the same as its height in the
current one, and its pointer is moved back by one unit, determining
which downstream neighbors receive grains.  One continues in this
fashion increasing the row $t$.

Unlike the DR model, eventually one
can encounter a site receiving three or more grains
from sites in the previous row.  If the total number of grains
received at a site, $n$, is even then the site
must topple exactly $n/2$ times.  The pointer at that site is moved back
$n/2$ steps, so $q_{i-1}(x,t)= q_i(x,t) -n/2$, reading the intervening
numbers in
the stack at that site to determine where the grains from that site
are sent.  If $n$ is odd and in
$C_i$ the height is one, then the site must have toppled $(n-1)/2$
times, with its height in $C_{i-1}$ being 0.  Thus $q_{i-1}(x,t)=
q_i(x,t) -(n-1)/2$. Similarly if $n$ is odd and in $C_i$ the height is
zero, then the site must have toppled $(n+1)/2$ times, with its height
in $C_{i-1}$ being 1.  Thus $q_{i-1}(x,t)= q_i(x,t) -(n+1)/2$.  One
reads the intervening sequence in the array of random numbers for that
site to determine how many grains each downstream neighbor receives,
and so forth.  Thus, given $C_i$, $x_i$, and $q_i(x,t)$, with a fixed
array $R(x,t,q)$, we can uniquely determine $C_{i-1}$ and
$q_{i-1}(x,t)$.  This proves the invertibility of the dynamics of
the SDM.

For a given array $R$ and set of pointers $\{q_i\}$, for any state $C_i$ there
are precisely $X$ distinct choices of $C_{i-1}$ and $C_{i+1}$
corresponding to the $X$ possible choices of $x_i$ and $x_{i+1}$.  It
then follows, as before, from the master equation for the evolution of
probabilities of configurations that the state prepared with a uniform
distribution over all stable states is invariant in time.  

Thus for the directed Manna model, the self-organized critical state is
the set of all stable states with equal likelihood; it is a product
measure state, where the probability for a site to be empty is equal
to the probability for it to have one grain, which are both equal to
1/2.  This is exactly the same as in the DR model, so for the SDM
the presence of multiple topplings does not lead to any correlations in
the states on the attractor.

\section{Discrete Equation for Avalanches in the Critical State}
 
The fact that the critical state is a product measure state
leads to a significant simplification;
namely the critical dynamics can be described as a type of generalized
branching process.  Thus one can simulate or describe avalanches
in an infinite system as follows.  
Consider a site which we we will call the origin.  
The origin in the equivalent branching process represents the site
that receives a grain in the critical state of the SDM.  The height at that
site is either one or zero with equal probability.  Add one grain to
it.  If the height now is greater than one it topples.  Then
define the heights at sites (1,1) and (-1,1); they are one or zero with
equal probability.  They receive grains from the origin according to
the stochastic rules of toppling in the directed model, and topple if
they are unstable.  In this way, one can always construct the lattice as the
avalanche propagates and one
can simulate the infinite system, albeit always for a finite time. 

We define the quantity $n(x,t)$ to be the number of grains added to
$(x,t)$ given that one grain was added to the origin.
The total number of grains that leave a site
$n_{out}(x,t)$ can at most differ by one
from the number of grains going in.  If $n(x,t)$ is even then
$n_{out}(x,t)=n(x,t)$.  If $n(x,t)$ is odd, then, since the number
of grains which  can leave any site is always even $n_{out}(x,t)= 
n(x,t) \pm 1$.  The process is critical and the increase or decrease
occur with equal probability.  { \it Thus we observe there is a source of
bounded, nonconservative noise, with a threshhold,
 in the dynamics of $n$ during an avalanche
that comes from the
presence or absence of grains in the metastable states.}
Since the number of grains going into a site can only arise as a consequence
of grains going into its immediate upstream neighbors we arrive at the
following discrete equation
\begin{eqnarray}
n(x,t+1) =  {1\over 2}\Bigl(n(x-1,t) + n(x+1,t)\Bigr) \qquad \qquad \qquad 
\nonumber \\
+ \theta_o(n(x+1,t))\eta(x+1,t) + \theta_o(n(x-1,t))\eta(x-1,t)  \nonumber \\
- -j(x+1,t) + j(x-1,t) \quad .   \nonumber \\
\end{eqnarray}
On average each site will get 1/2 of the grains going into its
upstream neighbors.  There are two sources of stochastic
variations from the average.
One is conservative:  Each upstream neighbor may divide its out flowing
grains unevenly between its two downstream sites, but what is taken
away from one downstream neighbor
is added to the other according to the binomial
distribution.   This gives a stochastic current $j$ which is either directed
to the right (here defined as positive) or to the left
(here defined as negative) for each site.
The first two moments of the stochastic current  of flowing
grains  are, from the binomial distribution,
\begin{eqnarray}
\langle j(x,t) \rangle & = & 0 \nonumber \\ 
\langle j(x,t)j(x',t')\rangle & = & {n(x,t)} 
\delta_o(x,x')\delta_o(t,t') \quad .  
\end{eqnarray}
Since this is a discrete equation, the Kronecker delta functions,
$\delta_o$,  are defined on the
set of integers. 

The nonconservative noise is the most interesting and, as we
shall see, relevant noise.  It is associated with the fact that the
metastable states either add or absorb flowing grains from the avalanche.
However, as mentioned before, the number of flowing grains can only
change by one unit irrespective of the local number of flowing grains as
long as it is nonzero.  This gives rise to the discrete Heaviside step
functions in Eq. ~2 defined as
$\theta_o(u) = 1$ for $u=1,2,3,\cdots$ and $\theta(u)=0$ otherwise.
With this convention the nonconservative noise is at each point in space-time
either $\pm 1$ with equal probability $1/4$ or 0 with probability $1/2$.  
Thus 
\begin{eqnarray}
\langle \eta(x,t) \rangle & = & 0 \nonumber \\ 
\langle \eta(x,t)\eta(x',t')\rangle & = &
{1\over 2}\delta_o(x,x')\delta_o(t,t') \quad .
\end{eqnarray}
The appropriate initial condition to describe the avalanche is
$n(x,0)=\delta_o(x,0)$.  The avalanche propagates and spreads out;
eventually it dies out.  Then a new avalanche, represented by a new
realization of the branching process is started.

\section{Continuum Equation for the Avalanches}

One could consider a rigorous derivation of the continuum limit of
Eqs. 2-4, taking the lattice size in space, $\Delta_x$,  and time
$\Delta_t$, as well as the grain size, $\Delta_n$, to zero.
Instead, here we invoke  the usual ``hand-waving'',
coarse graining procedure to obtain a smooth function $n$ of continuous
variables $x$ and $t$.  
Expanding to leading order in gradients,
and time derivatives, we arrive at
\begin{equation}
{\partial n(x,t) \over \partial t} = {1\over 2} \nabla^2 n(x,t) -
2{\partial j(x,t) \over \partial x} + 2\theta(n(x,t))\eta(x,t) \quad ,
\end{equation} where the threshold function
$\theta(u)=0$ for $u \leq 0$ and $\theta(u)=1$ for
$u > 0$.  By the central limit theorem, the noise terms are both
Gaussian with first and second moments
\begin{eqnarray}
\langle \eta(x,t) \rangle & = & 0 \nonumber \\ 
\langle \eta(x,t)\eta(x',t')\rangle & = &
{1\over 2}\delta(x-x')\delta(t-t') \nonumber \\
\langle j(x,t) \rangle & = & 0 \nonumber \\ \langle
j(x,t)j(x',t')\rangle & = & n(x,t) \
\delta(x-x')\delta(t-t') \quad .
\end{eqnarray}

The appropriate initial condition
for the avalanche is $n(x,0)= \delta(x)$.  The avalanche grows by increasing
or decreasing $n$ locally where $n$ is non-zero.  Eventually the avalanche
dies and $n(x,t)=0$ everywhere.  This equation describes the
transient out of an absorbing state associated with the avalanche.  In
particular, the state with no flowing grains $n(x,t)=0$ for all
$(x,t)$ is stable, which is a requirement of any equation describing
avalanche dynamics.

\subsection{Analysis}
Dimensional Analysis is the simplest tool we can apply, and the first
step in any theoretical analysis.
The dimension of the conservative noise is $[j]^2/[x]^2 = ([n]/[t][x]^3)$,
and dimension of the nonconservative noise is $[\eta]^2 =(1/[x][t])$.  
Thus, as long as $[n]< [x]^2$ then the
conservative noise is irrelevant with respect to the nonconservative noise.
Ignoring this term we arrive at our main result
\begin{equation}
{\partial n(x,t) \over \partial t} = {1\over 2} \nabla^2 n(x,t) +
2\theta(n(x,t))\eta(x,t) \quad .
\label{ew}
\end{equation}
In the region covered by the avalanche $n(x,t)>0$, the threshold
function, $\theta=1$ and may be ignored, resulting in an avalanche dynamics
described by the linear Edwards-Wilkinson equation \cite{ew}.  Dimensional
analysis then gives the correct scaling of various quantities.  Thus
for the SDM, $x_c \sim t_c^{1/z}$ with $z=2$ precisely as in the DR model.
However, the Edwards-Wilkinson equation gives a rough surface in one
dimension and the maximum number of topplings scales as the transverse
extent of the avalanche as $n_c \sim x_c^{1/2}$.  This differs markedly
from the DR model where $n_c = 1$, independent of the transverse extent,
$x_c$.

Continuing with our scaling analysis,
the area covered by the avalanche is $a \sim x_ct_c \sim t_c^{3/2}$
(as in the DR model), but the
size of the avalanche includes the extra effect of multiple topplings.
The size scales as $s \sim nx_ct_c \sim t_c^{7/4}$.  Since on average for
every grain added one grain must be transported the entire length of
the system to the open boundary, we have that $<s> = T$.  Since all
the geometric quantities associated with avalanches exhibit scaling,
it is reasonable to assume, and can perhaps be proven, that the
distribution of avalanche sizes, times, and spatial extent are power
laws, namely $P(s,T) \sim s^{-\tau}f(s/T^D)$, $P_t(t,T) \sim
t^{-\tau_t}g(t/T)$, and $P_x(x,T) \sim x^{-\tau_x}w(x/T^{1/z})$.  The
constraint on the average size then gives $1=D(2-\tau)$ or
$\tau=10/7$.  Similarly, from conservation of probability, $\tau_t -1
=D(\tau -1)=(\tau_x -1)/z$, gives $\tau_t=D=7/4$ and $\tau_x = 5/2$.

It is straightforward to check that Eqs. ~2-4 also apply to the
case where there are $d$ transverse dimensions.  The only factors that
are changed are various constants.  Applying dimensional analysis,
we find that in $d$ transverse dimensions, $x_{c} \sim t_c^{1/2}$,
so that $z=2$ and
$n_c \sim x^{2-d\over 2}$.  The upper critical dimension is $d_c=2$ above which
the maximum number of topplings does not diverge with the
size of the avalanche, and mean field results obtain.
 This corresponds to the fact that
the surface described by the EW equation is flat above two dimensions
rather than being rough.
For $d \leq 2$, $s \sim n_c x_c^d t \sim t^{3/2 + d/4}$, i.e. 
$D=3/2 + d/4$ and the other exponents are obtained via the above
scaling relations giving $\tau= 2 - (4/(6+d))$, and $\tau_t = 3/2 + d/4$.

\subsubsection{The Threshold Term}

Outside the region covered by the avalanche, the threshhold function
$\theta$
has a major effect on the dynamics.  In particular, in regions where
$n(x,t)=0$, the interface is pinned and cannot move.  The noise does
not act where there are no flowing grains! This is completely
different than the usual models of stochastic interfacial
growth.   The threshhold function
importantly breaks the translational symmetry of the EW equation 
($n \rightarrow n + \mbox{const}.$) and
leads to an absorbing state.  Typically absorbing state phase
transitions have been considered where the amplitude of the noise
depends on the activity $n$ to some positive power \cite{as}.  Here we find a
very weak effect simply distinguishing between having activity and not
having it in terms of a threshold function.  This effect is so weak
that the scaling dimensions of the propagating avalanche are the same
as the linear EW equation.  Obviously if the threshhold function
 $\theta(n)$ were replaced
by $n^{\alpha}$ in Eq. ~5   that would no longer be the case.  Thus
Eqs. ~5-6 is a hybrid combining  interface dynamics (the
number of topplings of the avalanche being the interface), and an
absorbing state model.  It describes a  previously undiscovered
 absorbing state phase transition.

\subsubsection{Averaging over Noise}

Averaging  over avalanches corresponds to averaging
Eq. ~7 over 
noise and we arrive at a linear diffusion equation for
the average propagation of flowing grains in response to a single
grain being added at $(0,0)$ to the critical system:
\begin{equation}
{\partial \langle n(x,t) \rangle \over \partial t} = {1\over 2} 
\nabla^2 \langle n(x,t) \rangle \quad ,
\end{equation}
whose solution is
\begin{equation}
\langle n(x,t) \rangle = {1\over (2\pi t)^{1/2}} e^{-x^2/2t} \quad .
\end{equation}
Obviously,
this solution has the important property of conservation, namely
$\int dx  \langle n(x,t) \rangle =1$ which is required for stationarity.
Note that the DR model also obeys exactly the same equation for
the average propagation of activity.  This equation is enforced by
the local conservation and symmetry properties of the system and is in no
way related to the presence or absence of multiple topplings.
Thus we get that for all models with the same symmetry and conservation
of grains, the dynamical exponent $z=2$ and  the exponent
$\eta=0$, since the average amount of activity remains constant in time.

\subsubsection{The Fokker-Planck Equation}
Ideally one would like to determine the full probability distribution
for the number of topplings $n(x,t)$ in  avalanches.  The
dynamics of this probability
distribution $P[n;t]$ is expressed by the Fokker-Planck equation.
The Fokker-Planck equation can be obtained by straightforward means from
the Langevin equation (Eqs. ~6,7).  It is 
\begin{eqnarray}
{\partial P[n;t] \over \partial t} & = &
- -{1\over 2}\int dx {\delta \over \delta n} \Bigl\{( \nabla^2 n)P\Bigr\}
\nonumber \\ & & + 
\int dx {\delta\over \delta n} \Biggl\{\theta(n)\Bigl\{
{\delta\over \delta n}(\theta(n)P)\Bigr\}\Biggr\} \quad .
\end{eqnarray}
Unfortunately we are not currently able to analyze this equation
in any significant way.

\section{Outlook for Future Work}
A major limitation of the present work is that applies only to a set
of directed models where all stable states are equally likely.  
Unlike most models of SOC, there are no spatial or temporal
correlations in the metastable states on the attractor.  Even in this
drastically simplified setting, the occurrence of multiple topplings
has a profound affect on the critical properties of the system,
changing the universality class.  This fact suggests that any
reasonable theory of avalanche dynamics in sandpile models of SOC must
treat the effect of multiple topplings.

Avalanches are described by the dynamics of particles which exhibit an
absorbing state phase transition.  As shown in our main result,
Eq. (7), this transition has the new feature that the amplitude of the
noise is a threshhold function of the local activity rather than a
being a power of the activity.  This threshhold form of the noise
amplitude has never before been discussed in the literature.  It may
be interesting to examine a broad range of absorbing state phase
transitions with such a threshhold noise amplitude, as well as the
implications of the threshhold noise amplitude on avalanche dynamics
in SOC.

The picture of avalanches as reaction-diffusion systems with an
absorbing state was first suggested in \cite{europhys} as applicable
to SOC systems and later in \cite{munoz}.  In the case discussed here,
the particles, representing topplings, are known to propagate in an
uncorrelated environment because the probability distribution of
metastable states on the attractor is described by a product measure.
In the general case, there will be important correlations from the
background that must be included along with boundary effects, leading
among other things, to correlations between avalanches.  It seems to
us that Dhar's development of an operator algebra for stochastic
models might provide a fruitful avenue to pursue further research.

{\it Note Added:} After  submitting this manuscript for
publication, another preprint appeared \cite{maslov}.
They also obtain the critical exponents for the stochastic
directed model we discuss.

We thank D. Dhar for helpful comments on the manuscript.  This work
was supported in part by NSF grant DMR-0074613.

\end{multicols}

\begin{references}
\bibitem{comment}This does not mean
that the operators of the stochastic model, i.e. the
``$a$'' operators in Ref. ~\cite{dharreview}, are invertible.
They are not invertible.
\end{references}
\end{document}